\RequirePackage{fix-cm}
\documentclass{svjour3}                     
\smartqed  
\usepackage{graphicx}
\usepackage{fix-cm}
\usepackage{ae}
\usepackage{enumitem}
\journalname{}
\begin{document}

\title{Quantized magnetic flux through the orbits of hydrogen-like atoms within the atomic model of Sommerfeld}

\author{W.-D. R. Stein
}


\institute{W.-D. R. Stein \at
              Helmholtz-Zentrum Berlin, Hahn-Meitner-Platz 1, D-14109 Berlin \\
              Tel.: +49-30-8062 43079\\
              Fax: +49-30-8062 43172\\
              \email{wolf-dieter.stein@helmholtz-berlin.de}           
}

\date{Received: date / Accepted: date}

\maketitle

\begin{abstract}
Within the Sommerfeld atomic model the quantization of magnetic flux through the electronic orbits is investigated together with its dependency on additional sources of magnetic fields. These sources alter the magnetic flux through the atomic orbits and in consequence are causing energy shifts. This effect is investigated for the cases, where the source is an external magnetic field, the magnetic moment of the nucleus or the magnetic moment of the electron. The energy shifts due to external magnetic fields, the magnetic dipole contribution of the hyperfine splitting and the spin-orbit coupling can be reproduced very well. The meaning of 'spin', however, changes within this approach drastically. The unusual Landé g-factor of 2 for the electron is a result of the orbital motion and the magnetic moment of the electron rather than it is an intrinsic property of the electron.
\end{abstract}

\section{Introduction}

The magnetic flux through the electronic orbits in the hydrogen atom was investigated by different methods within several atomic models, as there are: the Schrödinger model \cite{Saglam2007a,Saglam2007b}, the Dirac model \cite{Saglam2006a} and the Rutherford-Bohr model \cite{Stein2012a}, showing in particular, that the magnetic flux through these orbits is quantized and has a pronounced spin-dependency. The quantization of magnetic flux in units of $\Phi_0 = h/e$ was first recognized in the 1950s by London \cite{London1950a} and Onsager \cite{Onsager1953a} by considering a supercurrent circulating around a closed path. This quantization (in units of $\Phi_0/2$) was observed only ten years later by Doll and Näbauer \cite{Doll1961a} and, independently, by Deaver and Fairbank \cite{Deaver1961a} in experiments dealing with the torque on superconducting rings (cylinders) in external magnetic fields. 

One method, which was used for studying the magnetic flux through the electronic orbits within the Schrödinger and the Dirac model, uses the conversion of the area-integral of the magnetic induction into a time-integral over the cyclotron period \cite{Saglam2002c}. The source of the magnetic field was taken to be the magnetic moment of the nucleus (here proton) \cite{Saglam2007a}. In ref. \cite{Stein2012a} it was discussed, that this approach fails to predict the magnetic flux through the orbits within the helium (${}^4$He) ion He${}^+$. However, by using a time-integrated version of Faraday's law of induction it was shown, that in the point-particle picture of the Bohr model, the magnetic flux through each electronic orbit, that fulfills the Bohr-Sommerfeld-Wilson (BSW) quantization rule, is an integer multiple of the magnetic flux quantum ($h/e$). According to Faraday's law, a change of the magnetic flux through an area gives rise to an electromotive force on a charge along the boundary of that area. Considering a time-interval, where the magnetic flux is changed adiabatically from its initial to its final value, a charge on the boundary of the area will be accelerated from an initial to a final speed and momentum, respectively. Due to the time-integration only the initial and the final state need to be considered. In the case of electrons, the time-integrated version of Faraday's law together with magnetic flux quantization turns out to be equivalent to the BSW quantization rule. By considering the magnetic flux from the magnetic moment of the nucleus as a disturbance, an energy shift of $3/8$-times the experimetal value of the hyperfine splitting of the ground state of the hydrogen atom can be reproduced.

Here, this method is applied to the more complicated but still classical model of the Sommerfeld atom \cite{Sommerfeld1916a}. In the case of electrons, the time-integrated version of Faraday's law together with magnetic flux quantization is still equivalent to the BSW quantization rule in the case of elliptic orbits. The energy shifts due to small external magnetic fields and the magnetic moments of the nucleus as well as of the electron are investigated within the Sommerfeld model of the atom. These shifts can be shown to be in good agreement with the well-known energy shifts according to the Zeeman effect, Paschen-Back effect, the magnetic dipole contribution of the hyperfine coupling and the spin-orbit coupling.  

The paper is organized as follows: in the next section the formalism is applied to the elliptic orbits of the Sommerfeld model. Thereafter a section discusses small disturbances due to additional magnetic fields in a simplified version which, however, leads to a better understanding of the basic rules. Only within this section the electron is considered to have a magnetic moment, but no 'spin' angular momentum. In the following sections the effects of external magnetic fields and the magnetic moments of the nucleus are discussed without that restriction.

The understanding of these effects in the Bohr-Sommerfeld model could be crucial for the understanding of the magnetic flux quantization in the Schrödinger and the Dirac model. These probability density based models, however, would need information about the structure of the magnetic field and are therefore much more complicated to study than the point-particle models.

\section{On the magnetic flux through the area enclosed by the elliptic orbit of the electron in the Bohr-Sommerfeld Atom}

Closed electronic orbits fulfilling the Bohr-Sommerfeld-Wilson (BSW) quantization rule enclose a magnetic flux which is an integer multiple of the magnetic flux quantum ($ \Phi_0 = h/e$) \cite{Stein2012a}. The magnetic flux enclosed by the electronic orbit can be calculated by considering the adiabatic acceleration of the electron due to increase of the magnetic flux through its orbit by means of Faraday's law of induction. In contrast to the derivation within the Rutherford-Bohr model of the atom not only one quantum number fulfills the BSW quantization rule, but two and in the case of external fields three quantum numbers have to be considered. 

According to Faraday's law of induction, the time-derivative of the magnetic flux through a region $\Sigma$ is opposite to the electromotive force (EMF) along the boundary $\partial \Sigma$ of that region:
\begin{equation}\label{eq:faraday}
\oint_{\partial \Sigma } \vec{E} \cdot d\vec{s} =: \mbox{EMF} = -\frac{d}{d t} \int_\Sigma \vec{B} \cdot d \vec{A} = -\frac{d}{d t} \Phi,
\end{equation}
where $\Phi$ is the magnetic flux through $\Sigma$. By time-integration of this equation and assuming an adiabatic acceleration of the electron with initially vanishing momentum, only the integration boundaries need to be considered, for an electron giving rise to
\begin{equation}\label{eq:quantum-condition}
\oint_{\partial \Sigma } \vec{p} \cdot d\vec{s} = e \cdot (\Phi_f - \Phi_i),
\end{equation}
where $\Phi_i$ is the initial and $\Phi_f$ the final magnetic flux trough $\Sigma$.
The left hand side is quantized for closed orbits according to the BSW quantization rule, and so is the right hand side, which implies a quantization of the magnetic flux $\Phi_f$ through the region $\Sigma$ for vanishing initial magnetic flux $\Phi_i$. Postulating, that the magnetic flux through the orbits is still quantized in the case of non-vanishing initial magnetic fluxes $\Phi_i$, this equation becomes, using the quantization condition $\Phi_f = n h/e$ for the final magnetic flux
\begin{equation}
\oint_{\partial \Sigma } \vec{p} \cdot d\textbf{s} = n h - e \Phi_i.
\end{equation}
The difference to the BSW quantization rule can be understood by introducing the vector potential $\mathbf{A}$ for the magnetic field corresponding to the initial magnetic flux and replacing the expression of magnetic flux by its vector potential
\begin{equation}
\oint_{\partial \Sigma } \vec{p} \cdot d\vec{s} = n h - e \oint_{\partial \Sigma } \vec{A} \cdot d\vec{s},
\end{equation}
which reduces to the definition of the canonical momentum:
\begin{equation}
n h = \oint_{\partial \Sigma } (\vec{p} + e \vec{A}) \cdot d\vec{s} = \oint_{\partial \Sigma } \vec{p}_{can} \cdot d\vec{s}.
\end{equation}
Here, energy shifts due to small initial magnetic fields will be studied. In analogy to the derivation of the energy for elliptic orbits originally done by Sommerfeld \cite{Sommerfeld1916a,Sommerfeld1960a}, the energy in the case of small disturbances can be derived by replacing the two generalized momenta $J_\varphi$ and $J_r$ as
\begin{equation}
J_\varphi = \oint \frac{\partial S}{\partial \varphi} d \varphi = n_\varphi h \qquad \mbox{by} \qquad J_\varphi = \oint \frac{\partial S}{\partial \varphi} d \varphi = n_\varphi h - e \Phi_\varphi
\end{equation}
and
\begin{equation}
J_r = \oint \frac{\partial S}{\partial r} d r = n_r h \qquad \mbox{by} \qquad J_r = \oint \frac{\partial S}{\partial r} d r = n_r h - e \Phi_r,
\end{equation}
where $S$ is Hamilton's principal function and $\Phi_\varphi$ and $\Phi_r$ are the initial magnetic fluxes associated to the corresponding quantum numbers $n_\varphi$ and $n_r$, respectively.
For the energy we find by using $\Phi = \Phi_\varphi + \Phi_r$ 
\begin{equation}
W = -\frac{m_e Z^2 e^4}{8 \varepsilon_0^2}\frac{1}{(n h - e \Phi )^2} \approx -\frac{m_e Z^2 e^4}{ 4 \varepsilon_0^2 n^2 h^2}( 1 + \frac{e \Phi}{n h}),
\end{equation}
where the approximation holds in the case of low magnetic fluxes $\Phi$ compared to the magnetic flux quantum.
The gross structure is given by the Bohr energy levels and the finer structures can be considered by proper initial magnetic fluxes.
The energy shifts due to small initial magnetic fluxes are
\begin{equation}\label{eq:energy-approximation}
\Delta W  \approx  \frac{m_e Z^2 e^5}{ 4 \varepsilon^2_0 n^3 h^3}  \Phi = 2 R_\infty c \frac{e Z^2}{n^3} \Phi.
\end{equation}
When considering the geometry of the orbits, the magnetic fluxes corresponding to the different quantum numbers need to be considered individually.
The modifications due to the small disturbances can be taken into account, by replacing
\begin{equation}
n h \quad \mbox{by} \quad n h - e \Phi, \quad n_\varphi h \quad \mbox{by} \quad n_\varphi h - e \Phi_\varphi, \quad \mbox{and} \quad n_r h \quad \mbox{by} \quad n_r h - e \Phi_r.
\end{equation}
In the corresponding equations for the geometry of the ellipse the semi-major axis $a$ changes from
\begin{equation}
a = \frac{n^2}{Z} a_0 \qquad \mbox{to} \qquad a = \frac{(n - \frac{e \Phi}{h})^2}{Z} a_0
\end{equation}
and the semi-minor axis $b$ from
\begin{equation}
b = \frac{n n_\varphi}{Z} a_0 \qquad \mbox{to} \qquad b = \frac{(n - \frac{e \Phi}{h})(n_\varphi - \frac{e \Phi_\varphi}{h})}{Z} a_0,
\end{equation}
where it becomes clear, that for elliptic orbits an initial magnetic flux $\Phi_\varphi$ alters the geometry of the ellipse in a different way, than the initial magnetic flux $\Phi_r$.

\section{Spin without angular momentum }

The electron has two properties, which are not independent of each other: its magnetic moment and its angular momentum. However, within this section the angular momentum of the electron will be neglected for simplification, but not its magnetic moment. The modifications when regarding the angular momentum of the electron will be discussed in the following section. The Zeeman effect and the hyperfine interaction as well as the spin-orbit coupling can be understood in the flux-quantum picture, where the spin angular momenta of the electron and the atomic nucleus are neglected, but not their magnetic moments. The additional magnetic flux through the atomic orbits will be calculated and in a linear approximation the energy shifts due to the additional magnetic flux are deduced.

\subsection{External magnetic field (Zeeman effect)}

The Zeeman effect describes the interaction of a constant external magnetic field with the atom. For small magnetic fields, where no changes of the geometry of the atomic orbits have to be considered, the magnetic flux through the elliptic orbit is 
\begin{equation}
\Phi_Z = \pi a b B \cos{\alpha} = \pi \frac{n^3 n_\varphi}{Z^2} a_0^2 B \cos{\alpha},
\end{equation}
where $a = n^2 a_0 / Z$ and $b = n n_\varphi a_0 / Z$ are the semi-major and semi-minor axes, $\pi a b$ is the size of the ellipse and $\alpha$ is the angle between the normal vector of the orbital plane and the direction of the magnetic field $B$. For much higher magnetic fields the change of the geometry of the atomic orbit has to be considered. The energy-shift due to the external magnetic field is according to equation (\ref{eq:energy-approximation}):
\begin{equation}
\Delta W  \approx  \frac{m_e Z^2 e^5}{ 4 \varepsilon^2_0 n^3 h^3} \left( \pi \frac{n^3 n_\varphi}{Z^2} a_0^2 B \cos{\alpha} \right) = \mu_B n_\psi B,
\end{equation}
where $\mu_B$ is the Bohr magneton and $n_\psi = n_\varphi \cos{\alpha}$ the quantum number according to the Sommerfeld atomic model. When interpretating $n_\psi$ as the magnetic quantum number $m=n_\psi$ this equation describes the energy shift for the normal (semi-classical) Zeeman effect.

\subsection{Magnetic flux from a magnetic dipole in one of the focal points of an ellipse}

For the two following effects, the magnetic dipole contribution to the hyperfine splitting and the spin-orbit coupling, the magnetic flux through an ellipse, where a magnetic dipole is in one of its focal points, needs to be calculated analogously to a magnetic moment in the center of the circlular orbit of the Bohr model. A parametrization of the orbit is
\begin{equation}
r (\varphi) = \frac{p}{ 1 - \varepsilon \cos{\varphi}},
\end{equation}
where $p$ is the focal parameter.
Integration of the magnetic field of a magnetic dipole
\begin{equation}
B = \frac{\mu_0}{4 \pi} \frac{\mu_f}{r^3},
\end{equation}
where $\mu_f$ is the component of the magnetic dipole orthogonal to the orbital plane, outside the ellipse (but within the plane of the ellipse) gives the magnetic flux
\begin{equation}
\Phi_{out} = \int_0^{2\pi} \int_{r(\varphi)}^\infty B r dr d\varphi =  \frac{\mu_0}{4 \pi} \mu_f \int_0^{2\pi}  \frac{1}{r(\varphi)} d\varphi = \frac{\mu_0}{2} \frac{\mu_f}{p}.
\end{equation}
As magnetic flux lines are supposed to be closed, the magnetic flux through an infinite plane should be zero and the flux through the elliptic orbit is $\Phi_{in} = -\Phi_{out}$. For the geometry of the ellipse one finds
\begin{equation}
\Phi = \frac{\mu_0}{2} \frac{\mu_f}{p} = \frac{\mu_0}{2} \frac{\mu_f a}{b^2} = \frac{\mu_0}{2} \frac{Z \mu_f n^2 a_0}{n^2 n_\varphi^2 a_0^2} = \frac{\mu_0}{2} \frac{Z \mu_f }{ n_\varphi^2 a_0},
\end{equation}
where the semi-major $a$ and semi-minor $b$ axes and their expressions depending on the quantum numbers $n$ and $n_\varphi$ have been used instead of the focal parameter $p$.

\subsection{Hyperfine splitting}

The second effect, the magnetic dipole contribution to the hyperfine interaction, is taken into account by considering the magnetic moment $\vec{\mu}_c$ of the nucleus in one of the focal points of the elliptic orbit. The magnetic flux through the elliptic orbit is (see previous section):
\begin{equation}
\Phi_{hf} =  \frac{\mu_0}{2}\frac{a \mu_c }{b^2} \cos{\beta} =  \frac{\mu_0}{2}\frac{Z \mu_c }{n_\varphi^2 a_0} \cos{\beta}.
\end{equation}
For small changes in the magnetic flux, the linear approximation of the energy is sufficient:
\begin{equation}
\Delta W  \approx  \frac{m_e Z^2 e^5}{ 4 \varepsilon^2_0 n^3 h^3} \left( \frac{\mu_0}{2}\frac{Z \mu_c }{n_\varphi^2 a_0} \cos{\beta} \right) = -\alpha^2 Z^3 h R_\infty c  \, \frac{\mu_c \cos{\beta}}{n^3 n_\varphi^2}.
\end{equation}
The correct hyperfine interval for the $1s$ orbit can be found by considering two states, where the magnetic moment of the atomic nucleus is pointing first in a direction which makes an angle $\beta$ with the normal vector of the ellipse and second in the opposite direction, where $n_\varphi = 1/2$ and $\cos{\beta} = 2/3$ is assumed. A derivation of the angle between the direction of the magnetic moment and the normal vector of the elliptic plane will be discussed in the following section, as this can be attributed to the interplay of the different angular momentum contributions. The value $n_\varphi = 1/2$ would mean, that the ground state is defined by the quantum numbers $n_r = n_\varphi = 1/2$. 
By assuming, that both quantum numbers $n_\varphi$ and $n_r$ start from $1/2$ with steps of one, the gross structure, where only the sum of both quantum numbers enters, will be equivalent to the gross structure of the Rutherford-Bohr model. Also the Zeeman level splitting is not affected from this assumption as the differences in $n_\varphi$ are still considered to be integers.

\subsection{Spin-orbit coupling}

The third effect, the 'spin-orbit coupling' can be modelled by putting the magnetic moment of the electron in one of the focal points of the elliptic orbit of the electron. Similar to the hyperfine interaction the magnetic flux through the atomic orbit with $\mu_e = g_s \mu_B$ is
\begin{equation}
\Phi_{ls} =  \frac{\mu_0}{2}\frac{a g_s \mu_B }{b^2} \cos{\beta} =  \frac{\mu_0}{2}\frac{Z g_s \mu_B }{n_\varphi^2 a_0} \cos{\beta}.
\end{equation}
For small additional external magnetic flux, the linear approximation of the energy is sufficient:
\begin{equation}
\Delta W  \approx  \frac{m_e Z^2 e^5}{ 4 \varepsilon^2_0 n^3 h^3} \left( \frac{\mu_0}{2}\frac{Z g_s \mu_B}{n_\varphi^2 a_0} \cos{\beta} \right) = Z^3 \frac{\mu_0}{4 \pi} g_s \mu_B^2 \frac{2 \cos{\beta}}{n^3 a_0^3 n^2_\varphi}.
\end{equation}
The expression for the 'spin-orbit coupling' differs from the well known expression only by a factor $\frac{Z n_\varphi^2}{12 \cos{\beta}}$ for the $2p_{3/2}$ state of hydrogen. 

In the following section the three effects will be studied while taking the spin angular momentum into account.


\section{Interplay of different angular momenta} 

Instead of interpretating the energy shifts of atomic energy levels due to the Zeeman effect, Paschen-Back effect and the hyperfine splitting as the additional energy of a magnetic moment within a magnetic field, these effects are here considered to be the result of the quantization of the magnetic flux through the atomic orbit in the case of a non-vanishing magnetic background field. Within the Sommerfeld atomic model two contributions (orbital motion, 'spin') to the magnetic flux through the orbit of the atom will be considered. One results purely from the orbital motion of the electron and one is due to the magnetic moment of the electron. 
The atom is considered to be a symmetric top with non-precessing total angular momentum. The angular momentum axis and the principal axis are in general not parallel.

The following points need to be considered for the description of the above mentioned effects to be described within the flux quantum picture.

\begin{enumerate}[label=\emph{\alph*})]
\item {\bf Different behaviour of orbital and spin contribution:}
Within the Sommerfeld model the electronic orbits are ellipses and their sizes are defined by the quantum numbers $n_r$ and $n_\varphi$, the orientation in space is given by a third quantum number $n_\psi = n_\varphi \cos{\alpha}$, where $\alpha$ is the angle between the normal vector of the elliptic plane and the direction of an external magnetic field (It is assumed, that there is always at least a very small one.). Here two contributions will be distinguished. One contribution results from the motion of the electron around the nucleus (orbital contribution) and the associated quantities are labelled with the index $l$. This contribution can be described similar to the electron motion of the electron within the original Sommerfeld model. The other contribution results from the magnetic moment ('spin') of the electron, where the associated quantities are labelled with the index $s$. This contribution is not present in the original Sommerfeld model. It could be interpretated as a combination of the additional magnetic flux through the orbit due to the magnetic field of the magnetic moment of the electron and an orbital motion to stabilize the orbit. The quantum numbers for the spin contribution are $n_r^s = n_\varphi^s = 1/2$. The combined effect will be described by the total quantum numbers, given by $n = n^l + n^s$, $n_r = n_r^l + n_r^s$, $n_\varphi = n_\varphi^l + n_\varphi^s$ and so on. The two contributions behave independent of each other. The index $j$ will be used for the combination of the orbital and the spin contribution.

\item {\bf Size of the atomic orbit:} 
For magnetic flux calculations the size of the atomic orbits is needed. The orbits are elliptic in the Sommerfeld model with size $A$ depending on the two quantum numbers $n$ and $n_\varphi$:
\begin{equation}
A = \pi a b = \pi \frac{n^3n_\varphi}{Z^2} a_0^2,
\end{equation}
where $a$ is the semi-major and $b$ the semi-minor axis. Here a small modification is necessary:
Similar to the length of the angular momentum vectors in quantum mechanics, the length of the vector area (the size of the area) is assumed to be
\begin{equation}
|\vec{A}| = \pi \frac{n^3}{Z^2}\sqrt{n_\varphi(n_\varphi + 1)} a_0^2,
\end{equation}
where the quantum number $n_\varphi$ was replaced by $\sqrt{n_\varphi(n_\varphi + 1)}$ in the semi-classical model. A discussion of the reasons for the replacement is not intended. But in probability-density based models, this might be explained by the difference between mean average and maximum value of the radius of the orbit.

\item {\bf Projection of vector areas to other vector areas:}
It is necessary to determine the size of the projection of a vector area into the direction of another vector area $\vec{A_1}\cdot \frac{\vec{A_2}}{|\vec{A_2}|}$.
Here it will be done exemplary for the two vector areas $\vec{A_l}$ and $\vec{A_j}$. The vector product will be calculated from squaring the expression $\vec{A_l}= \vec{A_j}- \vec{A_s}$, which is equivalent to the postulation of a linear summation of vector areas:
\begin{equation}
\vec{A_l}\cdot \frac{\vec{A_j}}{|\vec{A_j}|} = \frac{\frac{1}{2}(|\vec{A_j}|^2 - |\vec{A_s}|^2 + |\vec{A_l}|^2)}{|\vec{A_j}|} 
\end{equation}
Inserting the sizes of the vector areas as described in $b)$ gives
\begin{equation}
\vec{A_l}\cdot \frac{\vec{A_j}}{|\vec{A_j}|} = \frac{\pi n^3 a_0^2}{2 Z^2} \; \frac{n_\varphi^j(n_\varphi^j + 1) - n_\varphi^s(n_\varphi^s + 1) + n_\varphi^l(n_\varphi^l + 1)}{ \sqrt{n_\varphi^j(n_\varphi^j + 1)}}.
\end{equation}
Analogously one finds for the projection of $\vec{A_s}$ into the direction of $\vec{A_j}$
\begin{equation}
\vec{A_s}\cdot \frac{\vec{A_j}}{|\vec{A_j}|} = \frac{\pi n^3 a_0^2}{2 Z^2} \; \frac{n_\varphi^j(n_\varphi^j + 1) + n_\varphi^s(n_\varphi^s + 1) - n_\varphi^l(n_\varphi^l + 1)}{ \sqrt{n_\varphi^j(n_\varphi^j + 1)}}.
\end{equation}

\item {\bf Projection of angular momenta to the direction of vector areas:}
In general, the angular momentum vector and the vector area are not parallel. Here an angle is proposed, resulting in a projection of the angular momentum in the direction of its corresponding vector area to be
\begin{equation}
\left(\frac{\vec{A_j}\cdot \vec{j}}{|\vec{A_j}|} \right) = n^j_\varphi \hbar = (n^l_\varphi \pm \frac{1}{2}) \hbar,
\end{equation}
where $n^j_\varphi = j$ and $n^l_\varphi = l$.

\item {\bf Magnetic flux through orbital plane from external magnetic fields:}
The magnetic flux $\Phi$ of a homogeneous magnetic field $\vec{B}$ through an orbital area with vector area $\vec{A}$ is
\begin{equation}
\vec{A} \cdot \vec{B} =  \pi \frac{n^3 n_\varphi}{Z^2} a_0^2 B \cos{\alpha},
\end{equation}
with $n_\varphi \cos{\alpha} = n_\psi$,
where the 'classical' size of the vector area (see $b$)) and the definition of Sommerfelds quantum number $n_\psi$ was used. Here, this is explained by the deviation of the vector area from the direction of angular momentum. An averaging effect occurs, resulting in a smaller value for the effective area seen from the magnetic field (similar to the rule $d$)).

\item {\bf Spin rule (g-factor):}
The orbital motion contribution caused by the spin of the electron has to be considered by postulating a spin rule. For the ground state already discussed, the the quatum numbers for the spin contributiuon are $n^s_\varphi = n^s_r = 1/2$. The anomalous gyromagnetic factor for the electron can be explained, by assuming, that the ratio between the radial and the orbital contribution remains always the same for the two spin quantum numbers and additional magnetic fluxes: 
\begin{equation}
n^s_\varphi = n^s_r, \qquad \mbox{and} \qquad \Phi^s_\varphi = \Phi^s_r.
\end{equation}
This condition makes sure, that in case of increasing magnetic flux $\Phi_\varphi$ the increase of the spin contribution $\Phi^s = \Phi^s_\varphi + \Phi^s_r$ is twice as large as other contributions not fulfilling the spin rule, like the orbital contribution. This assumption leads to a g-factor of 2.
\end{enumerate}


Using these rules, several effects are studied in following.

\subsection{Zeeman effect}

Because of spin-orbit coupling for weak external magnetic fields, the spin and the orbital part are not independent and only the projections of the spin vector area $\vec{A_s}$ and the orbital vector area $\vec{A_l}$ in the direction of the total vector area need to be considered. Keeping in mind the rule ($f)$) of the equivalence of the two spin quantum numbers $n_\varphi^s$ and $n_r^s$, a factor of 2 has to be applied in front of the area of the spin contribution, resulting in the additional magentic flux
\begin{equation}
\Phi_Z \propto (2 \vec{A}_s + \vec{A}_l) \cdot \vec{B}.
\end{equation}
Due to the coupling of the spin and the orbital contribution, the projection of these vector in the direction of the combined vector area $\vec{A}_j$ enter the equation of magnetic flux
\begin{equation}
\Phi_Z = A_{proj.} B_{proj.} = \frac{(2\vec{A_s} + \vec{A_l})\cdot \vec{A_j}}{|\vec{A_j}|} \frac{\vec{A_j} \cdot \vec{B}}{|\vec{A_j}|}.
\end{equation}
The projection of the vector areas to the direction of other vector areas are given in the previous section (see rule $c)$). Here only the case of weak magnetic fields is considered, where the deformation of the geometry is neglectable. The effective area hence is
\begin{equation}
\frac{(2\vec{A_s} +  \vec{A_l}) \cdot \vec{A_j}}{|\vec{A_j}|}  = \frac{\pi n^3 a_0^2}{2 Z^2} \; \frac{3 n_\varphi^j(n_\varphi^j + 1) + n_\varphi^s(n_\varphi^s + 1) - n_\varphi^l(n_\varphi^l + 1)} {\sqrt{n_\varphi^j(n_\varphi^j + 1)}}.
\end{equation}
However, the vector area is parallel to the axis of symmetry of the top, as which the atom is considered and not parallel to the direction of the angular momentum and hence rotating around the direction of the magnetic field. As the full angular momentum is assumed to be constant in space, the angular momentum of the nucleus and the angular momentum are circulating around the direction of the full angular momentum. The projection of the magnetic field vector $\vec{B}$ in the direction of the area vector $\vec{A}_j$ leads with (see rule $b)$)
\begin{equation}
\vec{A_j} \cdot \vec{B} =  \pi \frac{n^3 n_\varphi^j}{Z^2} a_0^2 B \cos{\alpha_j}, \; \mbox{and} \; |\vec{A_j}| = \pi \frac{n^3}{Z^2}\sqrt{n_\varphi^j(n_\varphi^j + 1)} a_0^2,
\end{equation}
to
\begin{equation}
\frac{\vec{A_j} \cdot \vec{B}}{|\vec{A_j}|} = \frac{n_\varphi^j \cos{\alpha_j} B}{\sqrt{n_\varphi^j(n_\varphi^j+1)}} = \frac{m_j B}{\sqrt{j(j+1)}},
\end{equation}
where $n_\psi^j$ and $n_\varphi^j$ have been identified by $m_j$ and $j$, respectively.
Combining these equations, the additional magnetic flux due to the external magnetic field is
\begin{equation}
\Phi_Z = \frac{\pi n^3 a_0^2}{Z^2} \underbrace{\left(1 + \frac{j (j+1) + s(s+1) - l(l+1)}{2j (j+1)}\right)}_{g_j} m_j B,
\end{equation} 
when identifying $n_\varphi^j$ with $j$, $n_\varphi^l$ with $l$ and $n_\varphi^s$ with $s$. The expression in the brackets is identical to the Landé factor $g_j$.
For the energy shifts one finds
\begin{equation}
\Delta W = \frac{m_e Z^2 e^4}{ 4 \varepsilon^2_0 n^3 h^3} e \Delta\Phi =  \frac{  m_e Z^2 e^4}{ 4 \varepsilon^2_0 n^3 h^3} e \pi \frac{n^3}{Z^2} a_0^2 g_j m_j B = \mu_B g_j m_j B.
\end{equation}
The Landé factor $g_j$ usually can be written in the form
\begin{equation}
g_j = 1 + \frac{j (j+1) + s(s+1) - l(l+1)}{2j (j+1)},
\end{equation}
by calculating the vector products of the angular momenta: $\vec{s} \cdot \vec{j}$ and $\vec{l} \cdot \vec{j}$.

\subsection{Paschen-Back-effect}

If the magnetic field is strong enough, the orbital angular momentum and the 'spin' angular momentum will not couple to a total angular momentum due to spin-orbit coupling as in the case of weak external magnetic fields, but will act independently. For the calculation of the magnetic flux, the time averaged vector areas for the orbital contribution $\vec{A}_l$ and for the spin contribution $\vec{A}_s$ need to be considered. Due to the equivalence of the the spin quantum numbers $n_\varphi^s$ and $n_r^s$ (see model property $f$)), a factor of two has to be considered for the spin contribution. The initial magnetic flux in the case of the Paschen-Back effect becomes
\begin{equation}
\Phi_{PB}  = ( 2 \vec{A}_s + \vec{A}_l ) \cdot  \vec{B}.
\end{equation}
The magnitude of the time averaged vector areas is proportional to the corresponding quantum numbers, resulting for the magnetic flux in (see rule b)):
\begin{equation}
\Phi_{PB} = ( 2 n^s_\varphi n^3 \frac{\pi a_0^2}{Z^2} \cos{\alpha_s} + n^l_\varphi n^3 \frac{\pi a_0^2}{Z^2} \cos{\alpha_l}) B,
\end{equation}
where $\alpha_s$ and $\alpha_l$ are the angles between the magnetic field and the vector areas of the spin and the angular momentum contribution, respectively. Using the quantization of the orientation in space $n_\psi = n_\varphi \cos{\alpha}$, the initial magnetic flux in the case of the Paschen-Back effects becomes
\begin{equation}
\Phi_{PB} =  ( 2 n^s_\psi + n^l_\psi) n^3 \frac{\pi a_0^2}{Z^2} B
\end{equation}
and the corresponding shift in energy with respect to the undisturbed orbit is
\begin{equation}
\Delta W_{PB} = \mu_B ( 2 n^s_\psi + n^l_\psi)B = \mu_B (2 m_s + m_l) B,
\end{equation}
where the quantum numbers $n^s_\psi$ and $n^l_\psi$ have been identified by the magnetic quantum numbers $m_s$ and $m_l$, respectively.

\subsection{Hyperfine interaction}

The hyperfine interaction will be described to be the change of energy resulting from the additional magnetic flux of the nucleus through the orbit of the electron due to the magnetic dipole. The magnetic flux through an elliptic orbit with focal parameter $p$ from a magnetic dipole $\mu_f$ orthogonal to the orbital plane in one of the focal points of the ellipse was shown to be
\begin{equation}
\Phi = \frac{\mu_0}{2} \frac{\mu_f}{p}.
\end{equation}
Simplified, the magnetic dipole is given by the projection of the magnetic moment of the nucleus $\vec{\mu}_I$ into the direction of the normal vector of the orbital plane $\frac{\vec{A_j} }{|\vec{A_j}|}$. The magnetic flux is
\begin{equation}\label{eq:hf-base}
\Phi_{hfs}^{simple} = \frac{\mu_0}{2} \frac{\vec{A_j} \cdot \vec{\mu}_I}{|\vec{A_j}|} \frac{1}{p} = \frac{\mu_0}{2} \frac{\vec{A_j}}{|\vec{A_j}|} \cdot \vec{\mu}_I \frac{a}{b^2} = \frac{\mu_0}{2} \frac{Z}{a_0} \frac{\vec{A_j} \cdot \vec{\mu}_I }{(n^j_\varphi)^2 |\vec{A_j}|} \quad \mbox{(simplified)},
\end{equation}
where the expressions of the semi-major and semi-minor axes have been used. 
However, the involved angular momenta, the spin angular momentum of the electron $\vec{s}$, the orbital angular momentum $\vec{l}$ and the angular momentum of the nucleus $\vec{I}$ define at the end the normal vectors of the different contributions and the direction of the magnetic moment of the nucleus. Expecting the time averaged normal vector of the electron orbit to be $\vec{j} = \vec{s} + \vec{l}$, both vectors, $\vec{A}_j$ and $\vec{\mu}_I$, will be replaced by the projection of each of these vectors into the direction of $\vec{j}$:
\begin{equation}
\Phi_{hfs} =   \frac{\mu_0}{2} \frac{Z}{(n_\varphi^j)^2 a_0} \left(\frac{\vec{A_j}}{|\vec{A_j}|} \cdot \frac{\vec{j}}{|\vec{j}|} \right) \frac{  \vec{j} \cdot \vec{\mu}_I }{ |\vec{j}|}.
\end{equation}
The projection of the angular momentum in the direction of the corresponding vector area was postulated in rule $d)$, and gives
\begin{equation}
\frac{\vec{A_j}\cdot \vec{j}}{(n_\varphi^j)^2 |\vec{A_j}|}   = \frac{(n_\varphi^l \pm \frac{1}{2})\hbar}{(n_\varphi^l \pm \frac{1}{2})^2} = \frac{g_s \hbar}{(2n_\varphi^l \pm 1)},
\end{equation}
where $g_s = 2$ and $n^j_\varphi = n_\varphi^l + n_\varphi^s$. With $\vec{\mu}_I = g_I \mu_K \vec{I}/\hbar$ and $\mu_0/(2 a_0 \hbar^2) = \pi \alpha^2/(2 m_e \mu_B^2)$ the additional magnetic flux through the electronic orbit is
\begin{equation}
\Phi_{hfs} = \frac{\mu_0}{2} \frac{Z}{a_0} \frac{g_s \hbar}{(2l \pm 1)} \frac{ g_I \mu_K \vec{j} \cdot \vec{I} }{ j(j+1)\hbar^3}=  \alpha^2 Z \frac{\pi}{2 m_e} \frac{g_s g_I \mu_K \vec{I} \cdot \vec{j}}{\mu_B^2 j (j+1)(2l \pm 1)}.
\end{equation}
With $\vec{I} \cdot \vec{j} = \hbar^2/2 [F(F+1)-I(I+1)-j(j+1)]$, $\mu_e = g_s/2 \frac{e \hbar}{2 m_e}$ and $\mu_{nuc} = g_I \mu_K I$ the additional magnetic flux caused by the magnetic dipole results in
\begin{equation}
\Phi_{hfs} =  \alpha^2 Z \frac{h}{2e} \frac{[F(F+1)-I(I+1)-j(j+1)]\mu_e \mu_{nuc}}{\mu_B^2 j (j+1)(2l \pm 1)I}.
\end{equation}
The energy shifts according to equation (\ref{eq:energy-approximation}) of the hyperfine levels amounts to 
\begin{equation}
\Delta W_{hfs} = 2 R_\infty c \frac{e Z^2}{n^3} \Delta\Phi_{hfs} =  \frac{A_{nlj}}{2}[F(F+1)-I(I+1)-j(j+1)]
\end{equation}
with
\begin{equation}
A_{nlj} = 2 \alpha^2  Z^3 R_\infty h c \frac{\mu_e \mu_{nuc}}{\mu_B^2 n^3 j (j+1)(2l \pm 1)I}.
\end{equation}
This expression differs from the usual expression for the hyperfine level shifts \cite{Kramida2010a}, when neglecting the reduced mass correction, the relativistic correction factor and the off-diagonal terms, only by the term $(2l \pm 1)$ which is in ref. \cite{Kramida2010a} as $(2l+1)$.

\section{Conclusions}

The quantization of magnetic flux through atomic orbits was investigated in more detail for the Sommerfeld atomic model. Neglecting the angular momentum of the constituents, effects like Zeeman effect, hyperfine splitting of atomic level and spin-orbit coupling can be explained approximately. Taking the angular momentum into account, Zeeman effect, Paschen-Back effect and hyperfine splitting of atomic level can be explained with high accuracy. The 'spin' needs to be seen from a different point of view. The unusual properties of the 'spin' are a result of the magnetic moment of the electron: The quantized magnetic flux through the orbit of the electron comes partly from the magnetic flux caused by the magnetic moment of the electron and partly from the orbital motion of the electron to stabilize the orbit, resulting in the g-factor of 2 for the electron. Rules have been proposed to explain the energy level shifts of several effects, which also contains corrections for the assumption of the electron to be a point-particle. It could be interesting to investigate a density based models, like Schrödinger equation and Dirac equation based models, with respect to energy shifts caused by additional magnetic fields. However, in this theories full vector field for the magnetic fields has to be considered.

\end{document}